\documentclass[final,english]{elsarticle}
\usepackage[T1]{fontenc}
\usepackage[latin9]{inputenc}
\usepackage{varioref}
\usepackage{units}
\usepackage{amsmath}
\usepackage{amssymb}
\usepackage{graphicx}
\usepackage{esint}
\usepackage{caption}
\usepackage{subcaption}
\pdfpageheight\paperheight
\pdfpagewidth\paperwidth

\journal{Elsevier}
\usepackage{upgreek}
\usepackage[bookmarks,bookmarksopen,bookmarksdepth=6]{hyperref}
\usepackage{cases}
\usepackage{url}
\usepackage{lineno}

\usepackage{a4wide}

\usepackage{newtxtext,newtxmath,amsmath}
\makeatother

\usepackage{babel}
\begin{document}


\title{SiPM understanding using simple Geiger-breakdown simulations}

 \author[]{R.~Klanner\corref{cor1}}
  \author[]{J.~Schwandt}

\cortext[cor1]{Corresponding author. Email address: Robert.Klanner@desy.de.
  Tel. +49~40~8998~2558.}

\address{ Institute for Experimental Physics, University of Hamburg,
 \\Luruper Chaussee 149, 22761, Hamburg, Germany.}


\begin{abstract}

 The results of a 1-D Monte-Carlo program, which simulates the avalanche multiplication in diodes with depths of the order of 1\,$\upmu$m based on the method proposed in Ref.\,\cite{Windischhofer:2023}, are presented.
  The aim of the study is to achieve a deeper understanding of silicon photo-multipliers.
 It is found that for a given over-voltage, $\mathit{OV}$, the maximum of the discharge current is reached at the breakdown voltage, $U_\mathit{bd}$, and that the avalanche stops when the voltage drops to approximately $U_\mathit{bd} - \mathit{OV}$.
  This is completely different to the generally accepted understanding of SiPMs, that the discharge stops at about $U_\mathit{bd}$.
 Simulated characteristics of the avalanche breakdown, like the time dependence of the avalanche current, over-voltage dependence of the Geiger-breakdown probability and the gain for photons and dark counts, are presented and compared to experimental findings.

\end{abstract}

\begin{keyword}
  Silicon photomultipliers \sep Geiger breakdown \sep simulations.
\end{keyword}

\maketitle
 \pagenumbering{arabic}

\section{Introduction}
 \label{sect:Introduction}

 SiPMs (Silicon Photo-Multipliers), matrices of avalanche photo-diodes, called pixels in this paper, operated above the breakdown voltage, are the photo-detectors of choice for many applications.
  They are robust, have single photon resolution, high photon-detection efficiency, operate at voltages below 100~V and are not affected by magnetic fields.
 In this paper an attempt is made to explain several of the experimentally observed properties of SiPMs with the help of a Monte Carlo simulation of the time dependence of the avalanche of a single pixel.

  \begin{figure}[!ht]
   \centering
    \includegraphics[width=0.5\textwidth]{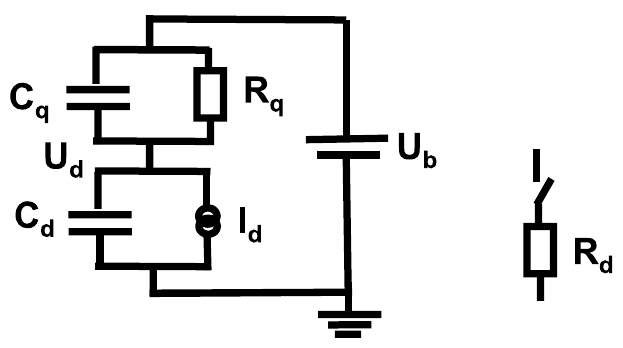}
   \caption{Left side: Electrical scheme of a single pixel of a SiPM.
    $U_b$ and $U_d$ are the bias voltage and the voltage over the avalanche region, respectively, $R_q$, $C_q$ and $C_d$ are the quenching resistor, the quenching capacitance and the capacitance of the avalanche region.
   A current source, $I_d(t)$, which describes the Geiger discharge, is in parallel to $C_d$.
    Right side: Example for a model of the current source.
   The switch closes at the start of the discharge and opens when $U_d $ reaches $U_\mathit{off}$, the voltage at which the Geiger discharge stops.}
  \label{fig:ElScheme}
 \end{figure}

 Fig.\,\ref{fig:ElScheme} shows an electrical diagram of a single pixel biased to the voltage $U_b$.
  The avalanche region has the capacitance $C_d$ and the current source generates the avalanche current, $I_d(t)$.
 The avalanche region is coupled to the power supply via the quenching resistor, $R_q$, which quenches the avalanche.
  During the avalanche, the voltage over the avalanche region, $U_d(t)$, decreases until
  the avalanche stops and there are no free charge carriers left in the avalanche region.
 Then, $U_d$ recovers to $U_b$ with the time constant $R_q \cdot (C_d + C_q)$, where the  quenching capacitance, $C_q$, which is parallel to $R_q$, produces a short current pulse across $C_d$.


 There is general agreement that this electrical scheme can be used to describe SiPMs.
  However, there are significantly different assumptions for both the time dependence of the discharge current, $I_d(t)$, as well as also for $U_\mathit{bd}$, the breakdown voltage above which avalanche breakdown occurs.
 In principle $U_\mathit{bd}$ can be calculated using the ionization integral as discussed in section\,\ref{subsect:Ubd}.
  However, neither the electric field distribution nor the ionization rates are sufficiently well known to yield reliable results.
 Therefore, $U_\mathit{bd}$ is determined from  current-voltage,  gain-voltage or  \emph{pde}-voltage (photon detection efficiency) measurements using phenomenological parameterizations.
  Although the results for $U_\mathit{bd}$ using these three methods are typically consistent within 0.1\,V, there are also difference of up to 1\,V.
 Several authors ascribe these differences to a difference between the voltage  $U_\mathit{bd}$ at which Geiger discharges are possible, and the voltage $U_\mathit{off}$, at which the Geiger discharges stop (Ref.\,\cite{Chmill:2016}).
  However, there is no consensus.

 For the discharge current $I_d(t)$ different ad hoc parameterizations can be found in the literature:
  One example is shown on the right side of Fig.\,\ref{fig:ElScheme}: A switch, which is open in the quiescent state, closes when the Geiger discharge starts, and opens again at the end of the Geiger discharge when the voltage $U_\mathit{off}$ is reached.
 In an other model the switch is connected to a discharge resistor, $R_d$, followed by a voltage source $U_\mathit{off}$. 
 Typically, $U_\mathit{off}$ is assumed to be equal to $U_\mathit{bd}$. 
 In both cases, the values of the components have to be chosen so that the duration of the discharge is less than 1\,ns.

  In this paper the time-dependent discharge current, $I_d(t)$, is simulated following the method of Ref.\,\cite{Windischhofer:2023}.
  In the following section the calculation of the breakdown voltage is discussed, followed by a presentation of the model used for the simulation of  avalanches.
  Using the simulation, the time dependence of the discharge current and of the over-voltage dependence of the discharge current, of the Geiger-discharge probability and of the gain for photons and for dark counts, are evaluated.
 This is followed by the simulation of more than one discharge channel.
  Finally, the main results are summarized and further tests of the used model of avalanche simulation suggested.

 \section{Simulation method and results}
 \label{sect:Method}

 In this section the simulation of the avalanche multiplication and Geiger breakdown is described, which follows the approach of Ref.\,\cite{Windischhofer:2023}.

  Simulated is a single square pixel of depth $d$ and area $d_\mathit{pix} ^{\,2} $, biased to the voltage $U_b$ and connected to the power supply by a quenching resistor $R_q$.
 The value of $R_q$ is sufficiently large, so that the current flowing through $R_q$ during the avalanche, which takes less than 1\,ns, can be ignored.
  The coordinate normal to the pixel surface is called $x$.
  The 1D electric field, $E(x)$, points in the $+\,x$\,direction, with the voltages $U(x=0) = 0$ and $U(x=d) = -U_b$.
  For the pixel capacitance $C_\mathit{d} = \varepsilon _0 \cdot \varepsilon _\mathrm{Si} \cdot d_\mathit{pix} ^{\,2} /d$ is assumed, and couplings to neighbouring pixels are ignored.

 \subsection{Breakdown voltage}
 \label{subsect:Ubd}

  First, the breakdown voltage, $U_\mathit{bd}$, as a function of $d$ and of the absolute temperature, $T$, is estimated by setting the Ionization Integral
  \begin{equation}\label{eq:Ionint}
    \mathit{II} = \int _0 ^d \alpha_e \exp \left(-\int_x ^d \left(\alpha_e - \alpha_h \right)\,\mathrm{d}x' \right)\mathrm{d}x =
    \frac{\alpha_e}{\alpha_e - \alpha_h} \Big(1 - \exp \big(-(\alpha_e - \alpha_h)\cdot d \big) \Big)
  \end{equation}
  equal to 1\,\cite{Sze:1981}.
   The hole- and electron-ionization coefficients, $\alpha_h$ and $\alpha_e$, which parameterize the mean number of electron-hole pairs generated for a 1\,cm path, are functions of the electric field, $E$, and of $T$.
  The right-hand side of Eq.\,\ref{eq:Ionint} gives the results for a constant $E$\,field.
   The hole and electron ionization coefficients in silicon are only poorly known\,\cite{Rivera:2022}.
  In addition, they do not take into account the "history" of the charge carriers responsible for the ionization, e.g. their momenta after the preceding interaction with the silicon lattice.
   This approach is a significant over-simplification, in particular for narrow amplification regions, as discussed in Ref.\,\cite{McIntyre:1999}.
   In this paper the van Overstraeten--de Man parametrization\,\cite{Overstraeten:1970} with the temperature dependence given in\,\cite{SYNOPSYS} is used, and it has been checked that other parameterizations give at least qualitatively similar results.

  \begin{figure}[!ht]
   \centering
   \begin{subfigure}[a]{0.54\textwidth}
    \includegraphics[width=\textwidth]{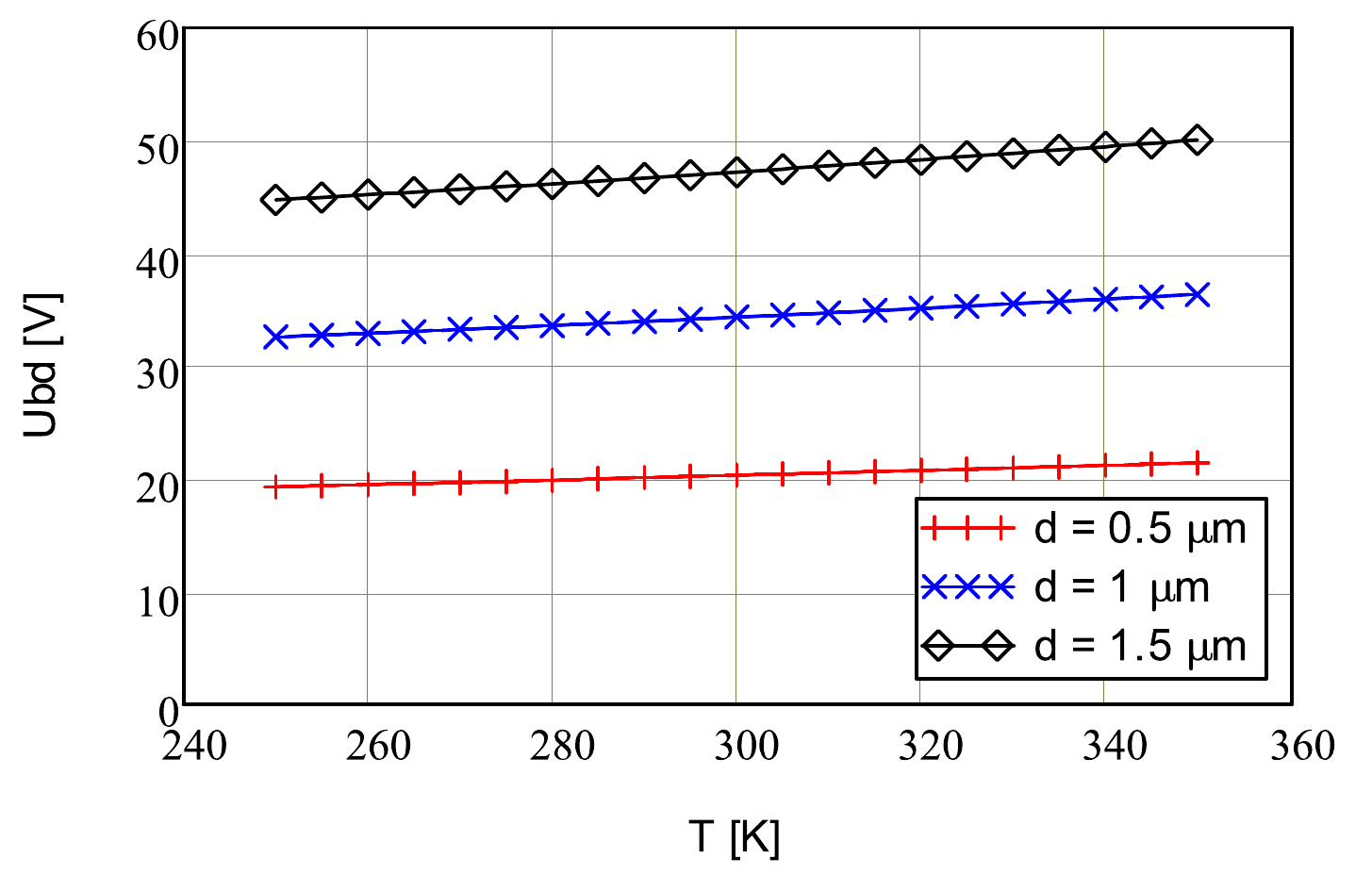}
    \caption{ }
    \label{fig:UbdT}
   \end{subfigure}%
    ~
   \begin{subfigure}[a]{0.48\textwidth}
    \includegraphics[width=\textwidth]{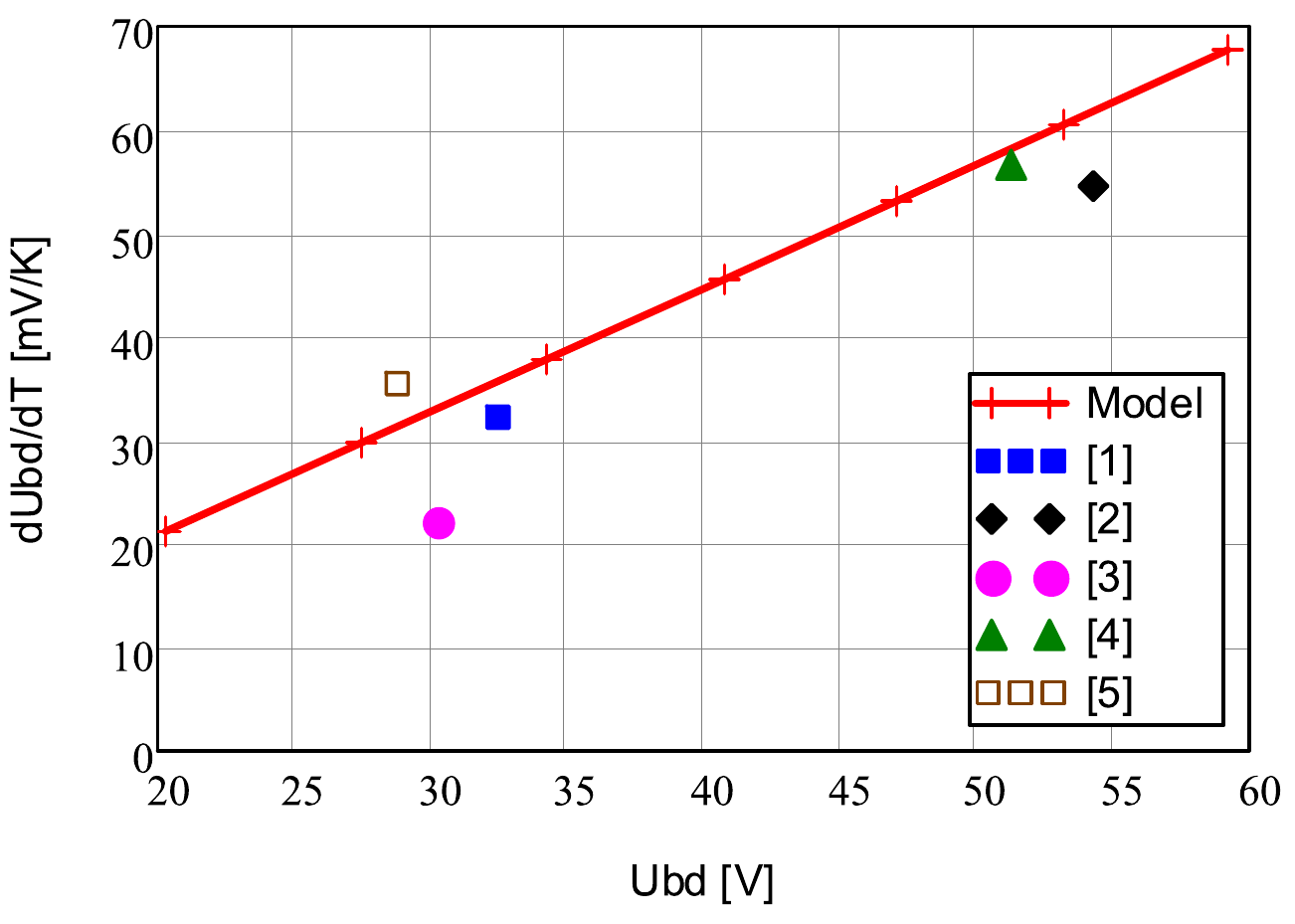}
    \caption{ }
    \label{fig:dUbddT}
  \end{subfigure}%
   \caption{ (a) Calculated breakdown voltage $U_\mathit{bd}$ for a constant electric field, $E$, as a function of the absolute temperature, $T$, for different depths $d$ of the multiplication layer, and
    (b) comparison of the calculated values of $\mathrm{d} U_\mathit{bd} /  \mathrm{d}\mathit{T}$ as a function of $U_\mathit{bd}$ to experimental results for the SiPMs
    [1] FBK NUV-HD, [2] Hamamatsu S13360-3050CS, and [3] SensL J-series 30035 (all Ref.\,\cite{Otte:2017}),
    [4] Hamamatsu S13161-3050AE-08 (Ref.\,\cite{Persiani:2023}), and
    [5] STMicroelectronics Catania R\&D, (Ref.\,\cite{Mazillo:2008}). }
  \label{fig:Ubd}
 \end{figure}

 Fig.\ref{fig:Ubd} shows some results of the calculated $T$ and $d$ dependence of $U_\mathit{bd}$.
  As also observed experimentally, $U_\mathit{bd}$ increases with $d$ and $T$, and if $U_\mathit{bd}$ is known the depth of the multiplication layer can be estimated.
 The calculations also confirm the observation that SiPMs with a higher $U_\mathit{bd}$ have a higher  $\mathrm{d} U_\mathit{bd} /  \mathrm{d}\mathit{T}$.
  In Fig.\,\ref{fig:dUbddT} the result of the model calculation is compared to experimental data.
 Although there is a significant spread of the experimental values at a given $U_\mathit{bd}$, the overall dependence of $\mathrm{d} U_\mathit{bd} /  \mathrm{d}\mathit{T}$ is reproduced.

  \subsection{Simulation of the avalanche multiplication}
 \label{subsect:Avalanche}

 In the most simple simulation, which is described next, a uniform electric field $E(t) = U_d(t)/d$ is assumed, with $U_d(t)$ the time dependent voltage drop over the avalanche region.
  Initially $U_d (0) = U_b$, the bias voltage.
 The depth $d$ is divided into $nx$ intervals of size $\Delta x = d/nx$.
  Given the high electric field, for both holes and electrons the saturation velocity $ v_\mathit{sat} = 10^7$\,cm/s (\cite{Sze:1981}) is assumed, which results in time steps $\Delta t = \Delta x / v_\mathit{sat}$ for charge carriers to drift from one grid point to its neighbour.
 This assumption results in a significant simplification of the simulation.
 Both, the impact ionisation and the induced charge can be considered as geometrical effects: The units of the ionisation coefficients and of the weighting field are 1/cm.
  Therefore, the integral of the current is not affected by this assumption, which has been verified by changing $v_\mathit{sat}$ by a factor two up and down.
   Only the time dependence of the induced current is affected, which is found to be inversely proportional to $ v_\mathit{sat}$.
 Typical values used in the simulation for $d = 1\,\upmu$m at $U_b = U_\mathit{bd} = 34.36 $\,V are: $E = 3.436 \times 10^5$\,V/cm, $nx = 100$, $\Delta x = 10$\,nm and $\Delta t = 10$\,ps.

 Next, the simulation of a single avalanche is discussed.
  Initially an electron-hole pair is generated at one $x$-grid point.
 Then the avalanche is iteratively calculated until there are no free charge carriers in the multiplication region.
  The number of holes at position $x_{ix}$ at the time step $it+1$, $nh_{ix}^{it + 1}$, and the corresponding number of electrons, $ne_{ix}^{it + 1}$, are obtained from the number of holes and electrons at the positions $x_\mathit{ix}$ at the time step \emph{it}

  \begin{equation}\label{eq:nht}
 nh_{ix}^{it + 1} = nh_{ix-1}^{it} + \mathrm{rpois}(nh_{ix-1}^{it} \cdot\alpha_h \cdot \Delta x) + \mathrm{rpois}(ne_{ix+1}^{it} \cdot \alpha_e \cdot \Delta x) \hspace{1mm},
  \end{equation}

  \begin{equation}\label{eq:net}
 ne_{ix}^{it + 1} = ne_{ix+1}^{it} + \mathrm{rpois}(ne_{ix+1}^{it} \cdot\alpha_e \cdot \Delta x) + \mathrm{rpois}(nh_{ix-1}^{it} \cdot \alpha_h \cdot \Delta x) \hspace{1mm},
  \end{equation}
 where rpois($\mu$) is a Poisson-distributed random number with mean $\mu$.

 Eq.\,\ref{eq:nht} can be understood in the following way:
  The first term describes the number of holes which drift by $+\,\Delta x$ during the time $\Delta t$, the second term the number of holes generated by the drifting holes, which also drift in the $+\,x$\,direction, and the third term the number of holes which are generated by the electrons which drift in the $-\,x$\,direction.
 This term assumes that the ionization by electrons takes place half way between the grid points.
  Similar arguments allow deriving Eq.\,\ref{eq:net} for electrons.
 To account for the charges flowing into the electrodes, $nh_{nx}$ and $ne_1$ are set to zero after every iteration.

 The avalanche current is obtained by the product of the total number of charge carriers, the elementary charge, $q_0$, the drift velocity, $v_\mathit{sat}$, and the weighting field, $1/d$:
 \begin{equation}\label{eq:Id}
   I_d(t) = \frac{q_0 \cdot v _\mathit{sat} \cdot \Sigma _\mathit{ix} (\mathit{nh}_\mathit{ix} + \mathit{ne}_\mathit{ix})}{d}.
 \end{equation}
 Using the model shown in Fig.\,\ref{fig:ElScheme}, the derivative, $\mathrm{d} U_d / \mathrm{d}t$, is obtained from current conservation:  The current in the quenching section $(U_b - U_d)/R_q + C_q \cdot \mathrm{d} (U_b - U_d) / \mathrm{d}t$ has to be equal to the current in the avalanche section $I_d + C_d \cdot \mathrm{d} U_b / \mathrm{d}t$, which gives
 \begin{equation}\label{eq:dUd}
   \frac{ \mathrm{d} U_d} {\mathrm{d}t} = \frac{(U_b - U_d)/R_q - I_d }{C_d + C_q} \approx - \frac{ I_d}{C_d},
 \end{equation}
 where the right most expression is the limit for $R_q \rightarrow \infty $ and $C_q = 0$, which is actually used in the simulations.
  It has been checked that this approximation is justified for realistic parameter values.
 Inserting Eq.\,\ref{eq:Id} into the right-hand side of Eq.\,\ref{eq:dUd}, and using for the time step of one iteration $\Delta t = \Delta x / v_\mathit{sat}$ gives
 \begin{equation}\label{eq:DeltaUd}
   \Delta Q^{it} = \frac{\left( \sum _{ix} (nh_{ix}^{it} + ne_{ix}^{it}) \right) \cdot q_0 \cdot \Delta x} {d} \hspace{5mm} \mathrm{and} \hspace{5mm}  U_d^{it+1} = U_d^{it} - \frac{\Delta Q^{it}}  {C_d},
 \end{equation}
 where $\Delta Q^{it}$ is the charge induced in iteration \emph{it} and $U_d^{it+1}$ the voltage over the avalanche region at step $\mathit{it} + 1$.
  It is noted that time does not enter in this relation, which means that the method is insensitive to the assumption on the drift velocities.
 For the electric field $E^\mathit{it} = U_d^\mathit{it}/ d$ is assumed; thus, space charge effects are ignored.

  \subsection{Time structure of avalanche development}
    \label{subsect:TimeAvalanche}

  In this section the time dependence of the number of charge carriers in the amplification region, which is proportional to the current, and of the voltage $U_d(t)$ are discussed.
   For the simulation an amplification region $d = 1\,\upmu$m and a constant electric field of strength $U_d(t)/d$ pointing in the $+\,x$\,direction are assumed.
  At time $t = 0$ an electron-hole pair is generated at $x = 0.99\,\upmu$m.
   The drift velocity of both electrons and holes is set to $v_\mathit{sat} = 10^7$\,cm/s which results in a transit time of 10\,ps.
  The calculated value of the breakdown voltage is 34.36\,V for the assumed temperature of 300\,K.

  \begin{figure}[!ht]
   \centering
   \begin{subfigure}[a]{0.46\textwidth}
    \includegraphics[width=\textwidth]{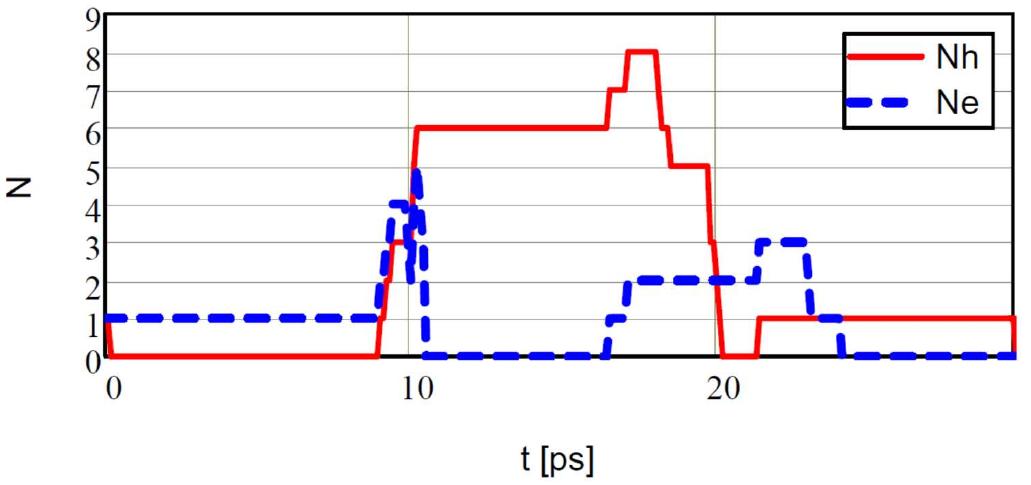}
    \caption{ }
    \label{fig:NtOV2nobd}
   \end{subfigure}%
    ~
   \begin{subfigure}[a]{0.54\textwidth}
    \includegraphics[width=\textwidth]{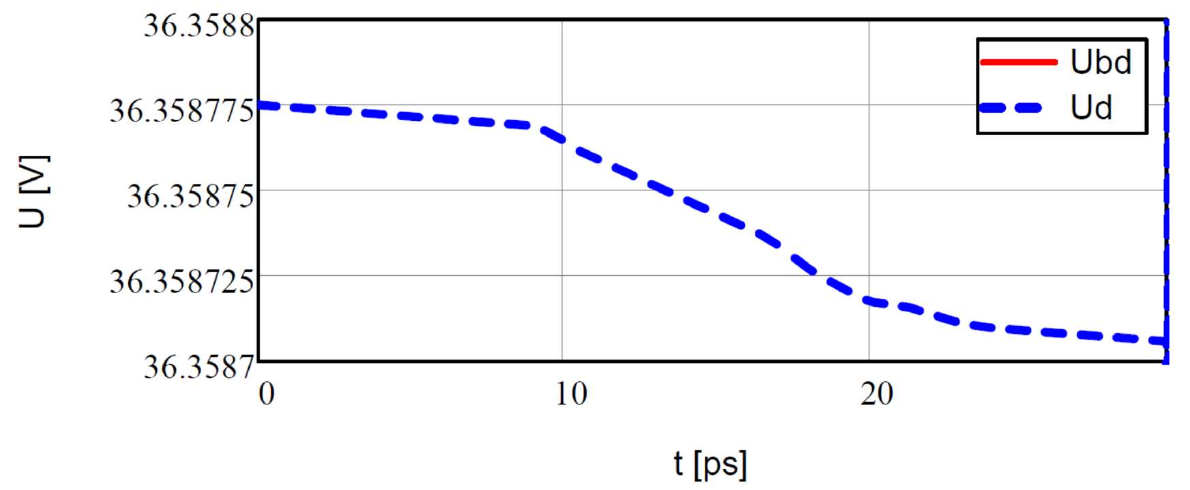}
    \caption{ }
    \label{fig:UdtOV2nobd}
  \end{subfigure}%

   \begin{subfigure}[a]{0.5\textwidth}
    \includegraphics[width=\textwidth]{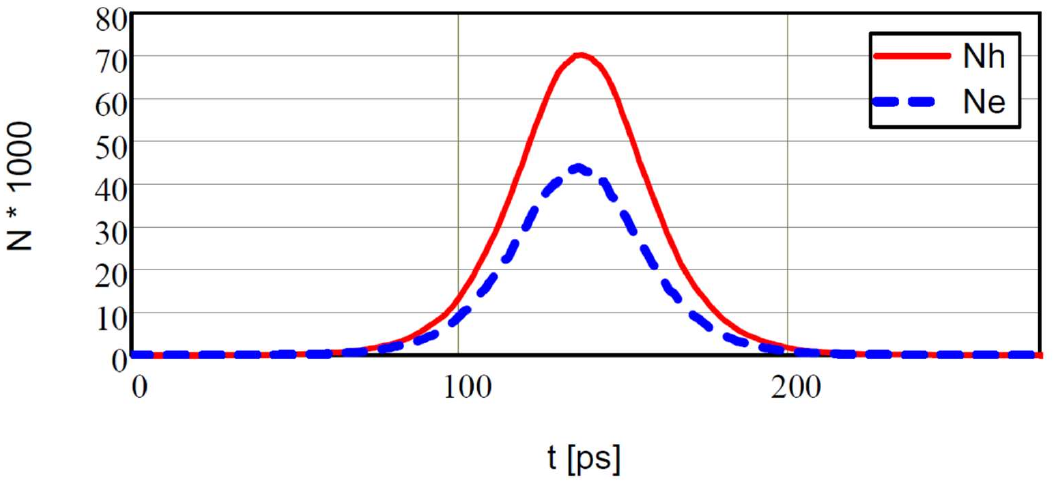}
    \caption{ }
    \label{fig:NtOV2bd}
   \end{subfigure}%
    ~
   \begin{subfigure}[a]{0.5\textwidth}
    \includegraphics[width=\textwidth]{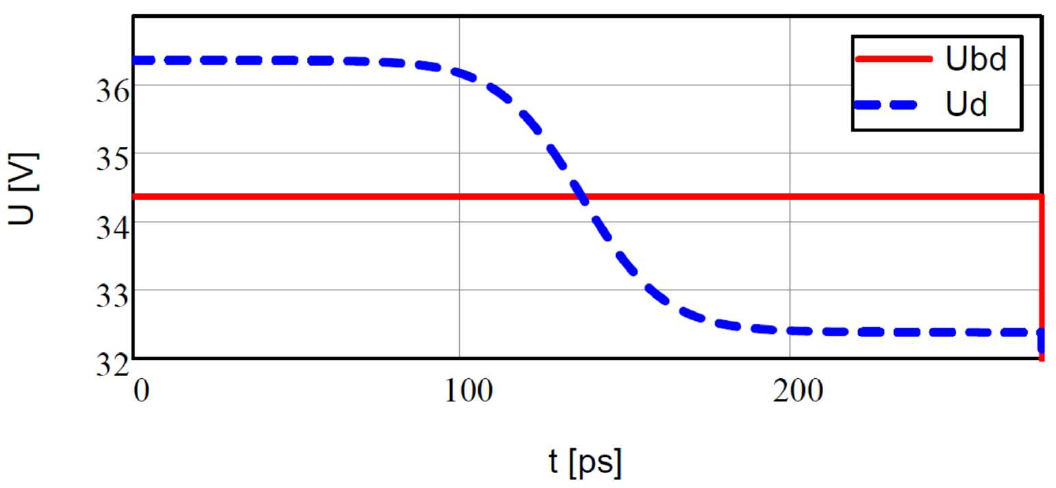}
    \caption{ }
    \label{fig:UdtOV2bd}
  \end{subfigure}%

   \begin{subfigure}[a]{0.5\textwidth}
    \includegraphics[width=\textwidth]{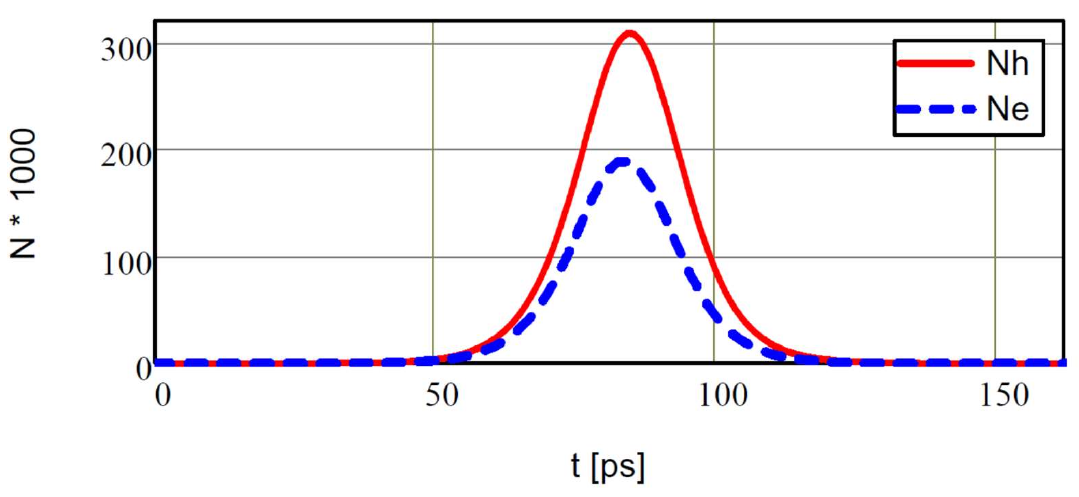}
    \caption{ }
    \label{fig:NtOV4bd}
   \end{subfigure}%
    ~
   \begin{subfigure}[a]{0.5\textwidth}
    \includegraphics[width=\textwidth]{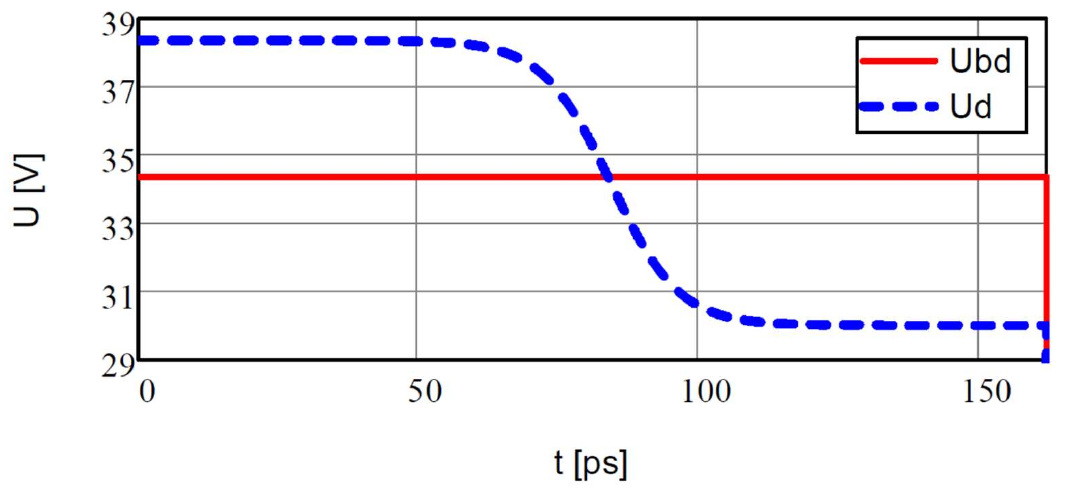}
    \caption{ }
    \label{fig:UdtOVbd}
  \end{subfigure}%
   \caption{ Time dependence of the number of charge carriers in the amplification region (left) and of the diode voltage (right) for
    $\mathit{OV} = 2$\,V without Geiger breakdown (top),
    $\mathit{OV} = 2$\,V with Geiger breakdown (middle), and
    $\mathit{OV} = 4$\,V with Geiger breakdown (bottom).
   In the simulation the depth of the amplification region $d = 1\,\upmu$m, an electron-hole pair is generated at $t = 0$ at $x = 0.99\,\upmu$m, the electric field points into the $+\,x$\,direction, and the temperature is 300\,K.}
  \label{fig:NUdt}
 \end{figure}

 Fig.\,\ref{fig:NUdt} shows for three simulated events the time dependence of the number of holes and electrons in the amplification region and the voltage $U_d (t)$ over the amplification region.
  For the event on top, the over-voltage $\mathit{OV} = 2\,V$ and no Geiger breakdown occurs.
 In Fig.\,\ref{fig:NtOV2nobd} one can see the occurrences of the ionizations and of the arrival of the charge carriers at the electrodes.
  At $t = 0$ the initial electron-hole pair is generated at $x = 0.99\,\upmu$m.
 The hole reaches the electrode after one time step of 0.1\,ps (solid line) and there are no holes and only a single electron (dashed line) drifts in the amplification region until the ionization at $t = 9$\,ps.
  The ionization at 9\,ps is followed by two close-by ionizations, so that shortly before the transit time of 10\,ps there are three holes and four electrons, all close to $x = 0$ in the amplification region.
 The drop of the number of electrons from four to two is caused by two electrons reaching the electrode at $x = 0$.
  Close to $t = 10$\,ps there are three more ionizations and the number of holes increases from three to six.
 As these ionizations occur close to $x = 0$ the electrons reach the electrode within about 1\,ps, and, until the ionization at 16\,ps, six holes and zero electrons drift in the amplification region.
  The further development of the number of charge carriers can be easily followed, until at $ t = 28$\,ps no more charge carriers are left.
 Including the primary ionization there are 10 ionizations.
  Thus the gain for this event $G = 10$, and the expected voltage drop over the pixel, using the relation $\Delta Q = G \cdot q_0 = C_d\cdot \Delta U$, is $68\,\upmu$V, where $C_\mathit{d} = 23.5$\,fF, the capacitance of a $15\,\upmu$m $\times \, 15\,\upmu$m pixel with a depth of $1\,\upmu$m, is used.
 This overall voltage drop is confirmed by the time dependence of $U_d(t)$ which is shown in Fig.\,\ref{fig:UdtOV2nobd}.

 Fig.\,\ref{fig:NtOV2bd} shows the time dependence of the number of charge carriers, $N(t)$, for an event with the same initial conditions as in the figure above.
   However, in this event a Geiger discharge has occurred.
  As discussed in detail in Ref.\,\cite{Windischhofer:2023}, after an initial time interval with large fluctuations there is an exponential increase of $N$ until $U_d \approx U_\mathit{bd}$, followed by an exponential decrease until $U_d \approx U_\mathit{bd} - \mathit{OV}$ is reached (see Fig.\,\ref{fig:UdtOV2bd}).
 The gain for this event is $5.8 \times 10^5$, the maximum of $N$ occurs at $t = 140$\,ps and the full width at half maximum is 35\,ps.
  The value of $U_d$ reached at the end of the Geiger discharge is 32.25\,V, which is 2.11\,V below $U_\mathit{bd} = 34.36$\,V.
 As already discussed in Ref.\,\cite{Windischhofer:2023}, the result of the simulation, that the Geiger discharge stops at $U_d \approx U_\mathit{bd} - \mathit{OV}$ and that the gain is $\approx 2 \cdot C_\mathit{d} \cdot \mathit{OV}$ which is completely different from the generally accepted understanding that the Geiger discharge stops at $U_d \approx U_\mathit{bd}$ and that the gain is $\approx C_\mathit{d} \cdot \mathit{OV}$, as discussed for example in Ref.\,\cite{Klanner:2019}.
  So far this issue has not been resolved.

\begin{figure}[!ht]
   \centering
    \includegraphics[width=0.7\textwidth]{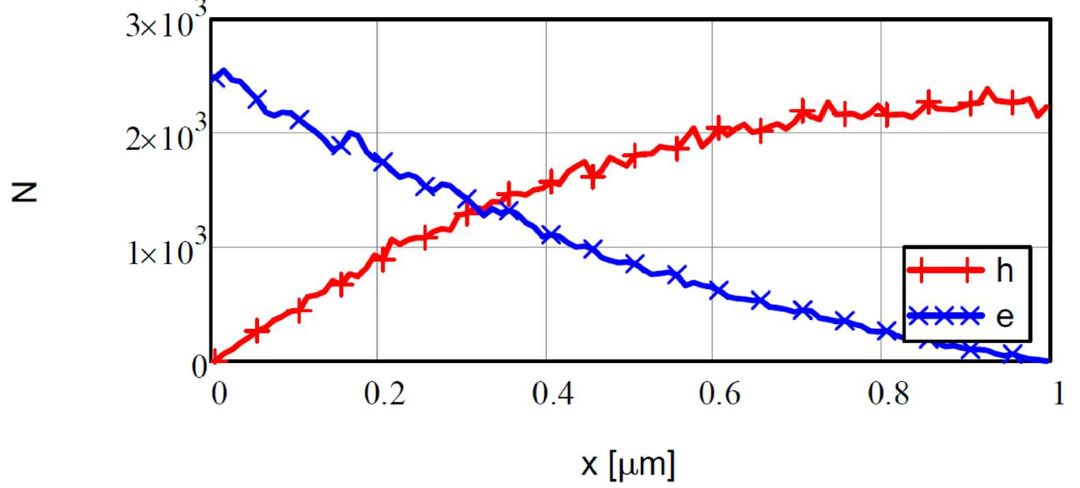}
   \caption{ $x$\,dependence of holes and electrons at the maximal current of the Geiger discharge.}
  \label{fig:Nx}
 \end{figure}

 It is also noted that holes contribute more to the total charge than electrons.
  The reason is that during the discharge the number of electrons, which have a higher ionization coefficient than holes, increases towards $x = 0$.
  Therefore, the number of ionizations is larger at smaller $x$, and  the mean path of the holes, which drift to $x = d$, is larger than for electrons.
 This is demonstrated in Fig.\,\ref{fig:Nx} which shows the $x$\,dependence of holes and electrons at the maximal current of the Geiger discharge.

  On the bottom of Fig.\,\ref{fig:NUdt} the results for a Geiger breakdown at $\mathit{OV} = 4$\,V are shown.
   Qualitatively, the time dependencies is similar to the breakdown at $\mathit{OV} = 2$\,V.
  However, the maximum of $N$ is at $t = 86$\,ps and the full width at half maximum is 23\,ps.
   The gain is $12.1 \times 10^5$, which is again $\approx 2 \cdot C_\mathit{d} \cdot \mathit{OV}$, and the value of $U_d$ reached at the end of the Geiger discharge is 30.02\,V.

  Simulations for different values of the depth $d$ of the depletion layer show that with increasing $d$ the gain decreases, because of the reduced pixel capacitance, and that both the time of the maximum of $N(t)$ and the full width at half maximum of $N(t)$ increase.

  One prediction of the simulation is that after a Geiger discharge $U_d(t) < U_\mathit{bd}$ for times $t < t_{1/2} = - \tau _s \cdot \ln (1/2)$ where $\tau _s = R_q \cdot C_d$ is the time constant of the recharging of the pixel.
   $R_q$ is the resistance of the quenching resistor and $C_d$ the capacitance of the avalanche region.
  Typical values for $\tau _s$ are in the range of 10\,ns to 100\,ns.
   As a consequence, after-pulses should appear only for $t > t_{1/2}$.
  Inspection of Refs\,\cite{Eckert:2010, Otte:2017, Piemonte:2019} shows that so far after-pulses have only been measured for $t$\,values greater than several tens of nanoseconds, and therefore this prediction of the simulation has not been tested so far.

  \subsection{Over-voltage dependence of the SiPM current, the breakdown probability and the gain}
    \label{subsect:OVdepPhotons}

  In this section basic characteristic of SiPMs, like the total signal charge, $Q_\mathit{tot}(OV)$, and the Geiger discharge probability, \emph{pG(OV)}, which can be compared to experimental results, are calculated for photons and dark counts.
   The simulations are for a pixel area of $15\,\upmu$m$\,\times\, 15\,\upmu$m, depths of the avalanche region of 0.5, 1.0 and 1.5\,$\upmu$m, and over-voltages between 0.5\,V and 10\,V.
  Photons with a wavelength of 350\,nm, which have an attenuation length of about 10\,nm in silicon, are simulated by generating the primary electron-hole pairs at $ x = 0.99\,\upmu$m, and dark counts by generating the primary electron-hole pairs uniformly over all $x$\,values.
 The over-voltage dependence, $Q_\mathit{tot}(\mathit{OV})$, is closely related to the voltage dependence of the SiPM current:
  If the rate of photons, $R_\gamma (\mathit{OV})$, and the contribution from cross-talk and after-pulses, \emph{XT(OV)}, are known, the photo-current $I_\gamma$ can be calculated by multiplying $Q_\mathit{tot}$ simulated for photons with $ R_\gamma \cdot (1 + \mathit{XT})$; if $I_\gamma(\mathit{OV})$ has been measured, \emph{XT(OV)} can be estimated.
 Similar relations hold for the dark-current, $I_\mathit{dark}$.

\begin{figure}[!ht]
   \centering
   \begin{subfigure}[a]{0.5\textwidth}
    \includegraphics[width=\textwidth]{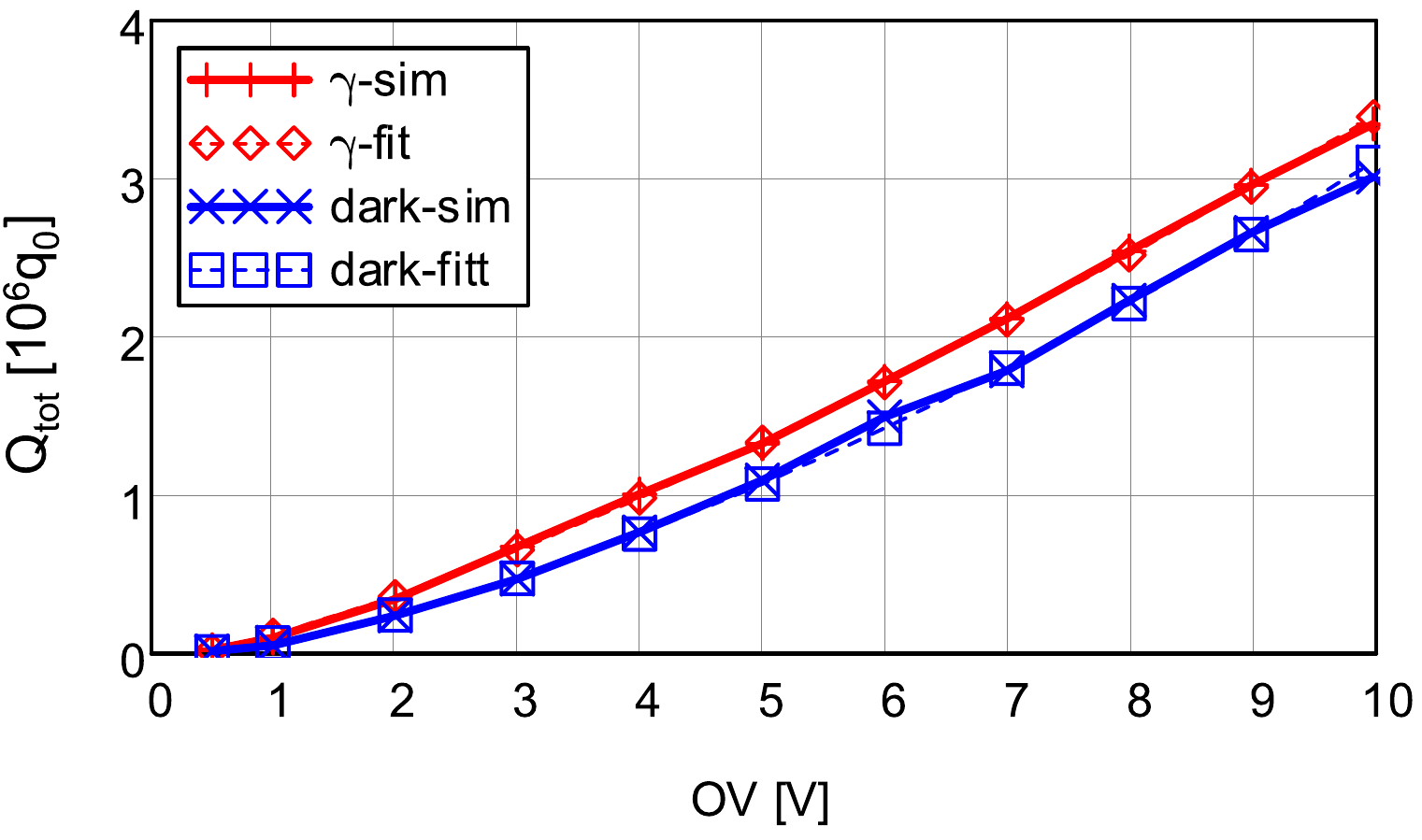}
    \caption{ }
    \label{fig:Iphd1}
   \end{subfigure}%
    ~
   \begin{subfigure}[a]{0.52\textwidth}
    \includegraphics[width=\textwidth]{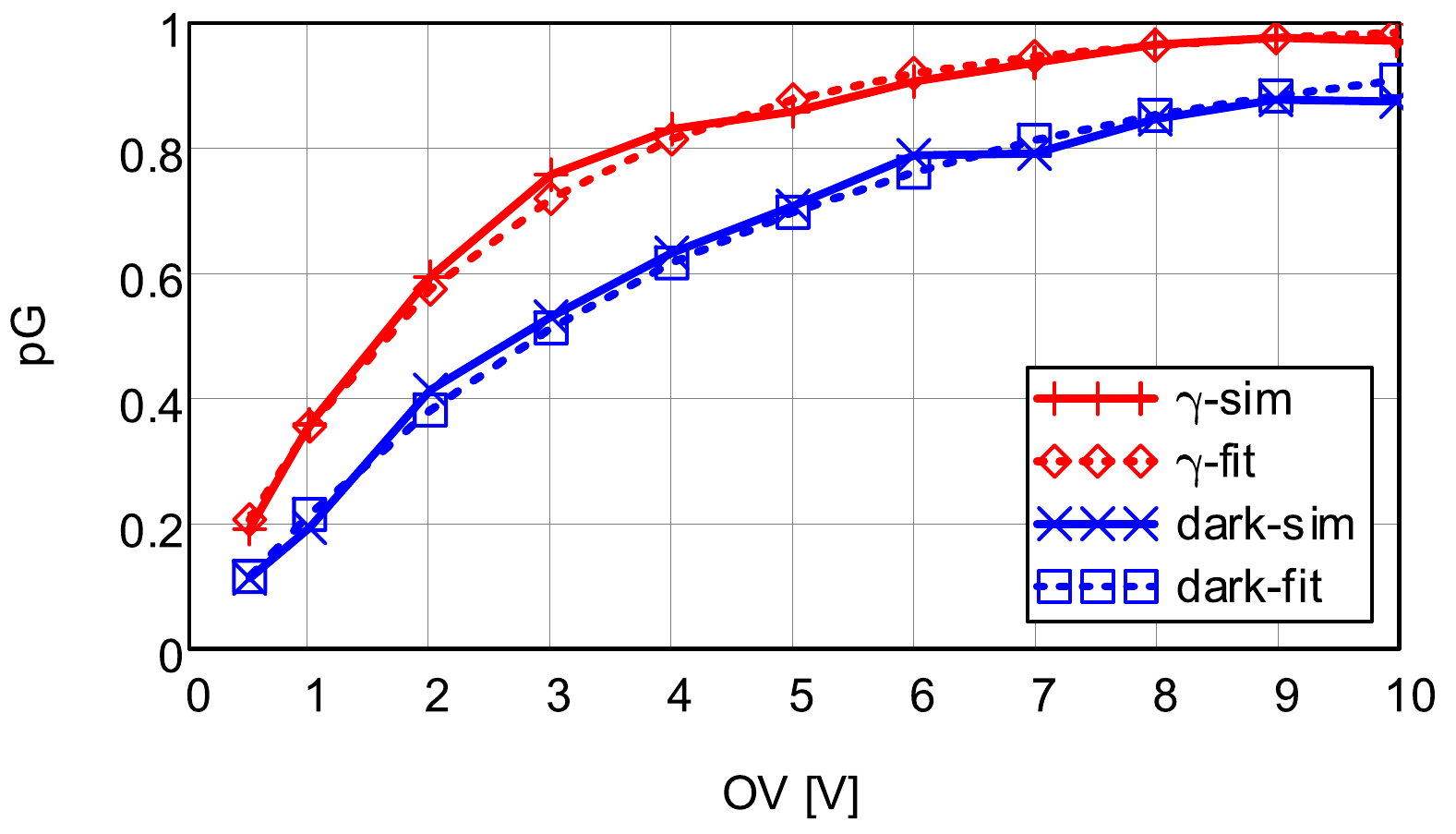}
    \caption{ }
    \label{fig:pGd1}
  \end{subfigure}%
   \caption{ (a) Simulated mean total charge, which includes events without and with Geiger discharges, as a function of the over-voltage for an avalanche region of $15 \times 15 \times 1 \, \upmu$m$^3$.
    The diamonds indicate the fit by Eq.\,\ref{eq:Iphd1} for photons ($\gamma$) and the sqares for dark counts (\emph{dark}).
    (b) Over-voltage dependence of the Geiger-breakdown probability for the same data, with fits by Eq.\,\ref{eq:pG1}. }
  \label{fig:pG1}
 \end{figure}

   For every over-voltage 1000 events are simulated.
  Fig.\,\ref{fig:Iphd1} shows to total SiPM charge, $Q_\mathit{tot}(\mathit{OV})$.
  The simulations include events without and with Geiger discharges.
 The voltage dependencies can be fitted by the function
 \begin{equation}\label{eq:Iphd1}
   Q_\mathit{tot} = \max\left(Q_\mathit{OV1} \cdot (\mathit{OV} - \Delta U_\mathit{tot})^a,0\right),
 \end{equation}
  where $\Delta U_\mathit{tot}$ is the over-voltage of the intercept of the fitted curve with the abscissa, $Q_\mathit{OV1}$ is the mean charge for $\mathit{OV} - \Delta U_\mathit{tot} = 1$\,V, and $a$ characterizes the increase of $Q_\mathit{tot} $ with $\mathit{OV}$.
 The values of the fitted parameters for the simulations with $d = 0.5$, 1 and 1.5\,$\upmu$m are shown in Table\,\ref{tab:dFit}.
  In all cases the fits describe the data within their statistical uncertainties.

  \begin{table} [!ht]
   \caption{Values of the parameters from the fits of Eqs\,\ref{eq:Iphd1} and \ref{eq:pG1} to the simulated data for different depths of the avalanche region, $d$, and the values of $U_\mathit{bd}$ from section\,\ref{subsect:Avalanche}.
    Rows 2 to 4 show the results for photons and rows 5 to 7 for dark counts.}
  \label{tab:dFit}
  \centering
  \begin{tabular}{c|c||c|c||c|c|c}
    \hline
    $d\,[\upmu$m] & $U_\mathit{bd}$ [V] & $\Delta U_\mathit{pG}$ [mV] & $U_0$ [V] & $Q_\mathit{OV1}\,[q_0]$ & $\Delta U_\mathit{tot}$ [mV] & $a$ \\
  \hline
    0.5 & $20.340 \pm 0.005$ &$-119 \pm 42$ & $2.19 \pm 0.06 $ & $(3.7\pm 0.3)\times 10^5$ & $256 \pm 110 $ & $1.29 \pm 0.03 $  \\
    1.0 & $34.359 \pm 0.007$ &$-56  \pm 60$ & $2.40 \pm 0.09 $ & $(1.8\pm 0.2)\times 10^5$ & $310 \pm 120 $ & $1.28 \pm 0.03 $  \\
    1.5 & $47.232 \pm 0.010$ &$ -42 \pm 41$ & $2.70 \pm 0.07 $ & $(1.1\pm 0.1)\times 10^5$ & $310 \pm 110 $ & $1.31 \pm 0.03 $  \\
  \hline
   0.5 & $20.340 \pm 0.005$ & $-88 \pm 46$ & $3.39 \pm 0.07 $ & $(2.5\pm 0.3)\times 10^5$ & $ 236 \pm 290 $ & $1.44 \pm 0.03 $  \\
   1.0 & $34.359 \pm 0.007$ & $-15 \pm 90$ & $4.19 \pm 0.14 $ & $(1.1\pm 0.3)\times 10^5$ & $ 280 \pm 320 $ & $1.47 \pm 0.09$ \\
   1.5 & $47.232 \pm 0.010$ & $ 52 \pm 60$ & $5.14 \pm 0.10 $ & $(0.5\pm 0.1)\times 10^5$ & $ 240 \pm 270 $ & $1.58 \pm 0.08$\\
  \end{tabular}
 \end{table}


 Fig.\,\ref{fig:Iphd1} shows that $Q_\mathit{tot}(\mathit{OV})$, i.e. the integral of the current per initial electron-hole pair, is larger for photons than for dark counts.
  The reason is that for the structure simulated, the electron path decreases
  for electron-hole pairs generated further away from the position at which the photons enter.
 As the ionization probability for electrons is larger than for holes, the decrease in electron-path length results in a smaller overall ionization probability.
   Whereas 350\,nm photons convert within about 10\,nm, the dark counts are approximately uniformly distributed in the amplification region.
 The values for $a$, given in Table\,\ref{tab:dFit}, show that voltage dependence of $I_\mathit{dark}$ is steeper than for $I_\gamma $, which is also observed experimentally.
  The simulated $a$\,values have no significant dependence on the breakdown voltage.
 For $\Delta U_\mathit{tot}$ a value of $\approx 300$\,mV is found, which is about three standard deviations larger than zero for all simulations.
  It is noted that the value of $U_\mathit{bd}$ obtained from the ionization integral may be different from the values in the literature, which are experimentally determined using different prescriptions, as discussed e.g. in Refs.\,\cite{Chmill:2016, Klanner:2019}.

 Events without and with Geiger discharges are well separated both in charge and in the time of the maximum of the discharge current.
  From the ratio of the number of events with a Geiger discharge to the total number of simulated event, the Geiger-discharge probability, \emph{pG(OV)}, is determined.
 Fig.\,\ref{fig:pGd1} shows the results for  $d = 1\,\upmu$m for photon- and dark-count-induced events together with fits by

 \begin{equation}\label{eq:pG1}
   \mathit{pG}(\mathit{OV}) = \max(1 - e^{-\left(\mathit{OV} - \Delta U_\mathit{pG}\right)/U_0}, 0),
 \end{equation}
 where $\Delta U_\mathit{pG}$ is the over-voltage of the intercept of the fitted curve with the abscissa, and $U_0$ the characteristic voltage of \emph{pG(OV)}.
  The values of the fitted parameters for the simulations with $d = 0.5$, 1 and 1.5\,$\upmu$m are shown in Table\,\ref{tab:dFit}.
 In all cases the fits describe the simulated data within their statistical uncertainties.
  The same parametrisation also describes experimental data (\cite{Otte:2017}) with similar values of the parameters.
   In Ref.\,\cite{Riegler:2021} an analytic equation for \emph{pG(OV)} for constant electric fields is presented.

 As expected, the Geiger-breakdown probability, $pG$, shown in Fig.\,\ref{fig:pGd1} for $d = 1\,\upmu$m, is significantly larger for photons than for dark-counts:
  At $\mathit{OV} = 0.5$\,V it is by a factor 2 bigger, and at $\mathit{OV} = 10$\,V the values are 98.5\,\% and 87.5\,\%, respectively.
 The values of $U_0$ reported in Table\,\ref{tab:dFit}, which describe the increase of \emph{pG} with \emph{OV}, increase with $U_\mathit{bd}$ and are significantly larger for dark-counts than for photons.
  Similar values and a similar increase have been reported by several authors, e.g.\,in Refs.\,\cite{Otte:2017, Rolph:2023}.
 The values of $\Delta U_\mathit{pG}$ are small and within their fairly large statistical uncertainties compatible with zero. It can be concluded, that the breakdown voltage can be determined from a fit to \emph{pG(OV)}.

  \begin{figure}[!ht]
   \centering
   \begin{subfigure}[a]{0.5\textwidth}
    \includegraphics[width=\textwidth]{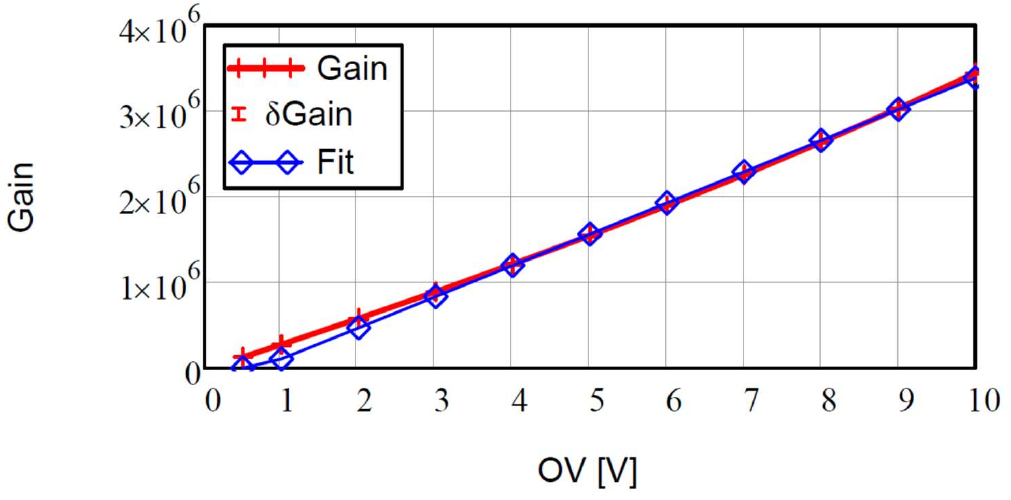}
    \caption{ }
    \label{fig:Gaind1}
   \end{subfigure}%
    ~
   \begin{subfigure}[a]{0.45\textwidth}
    \includegraphics[width=\textwidth]{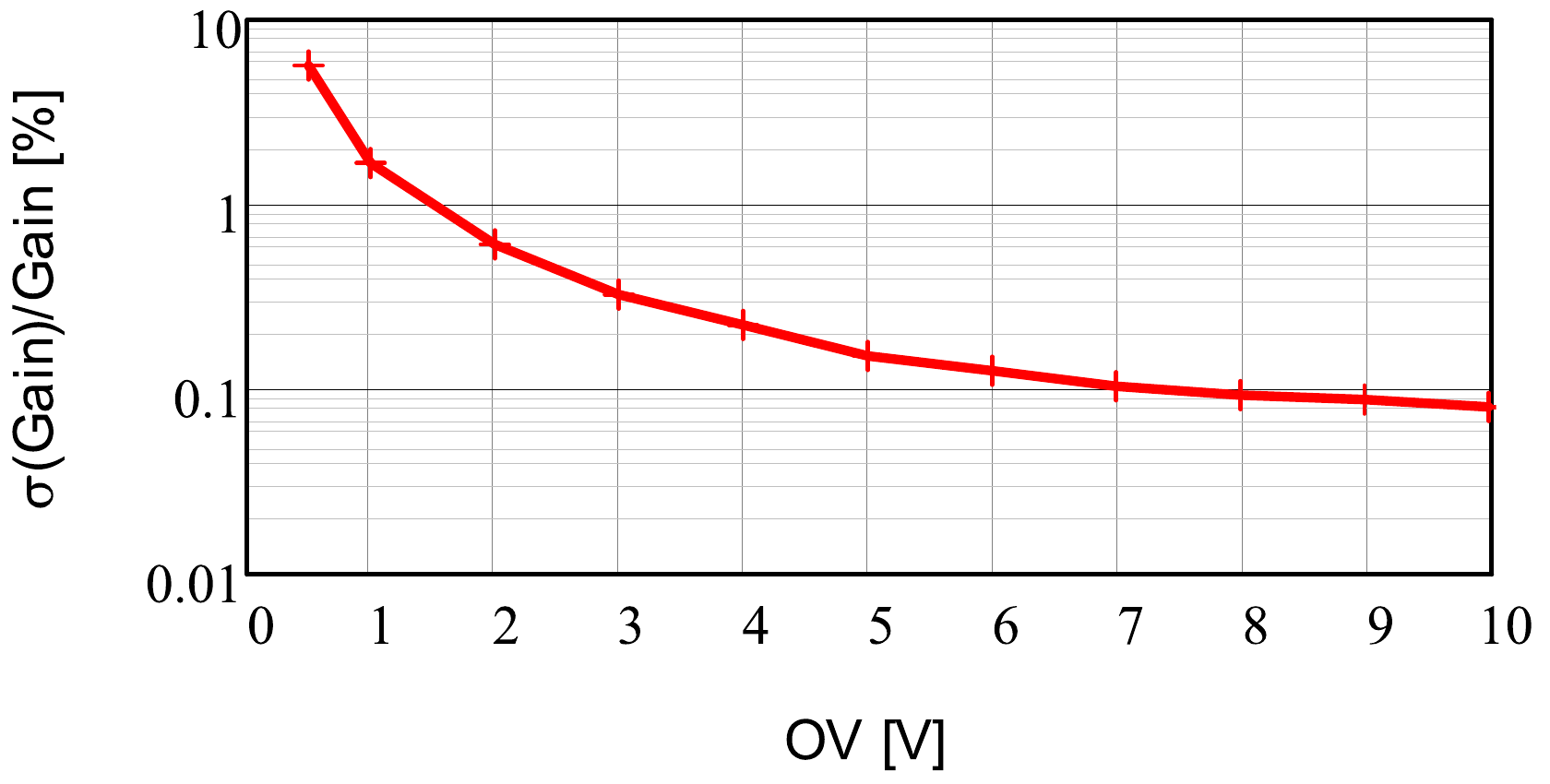}
    \caption{ }
    \label{fig:SigGaind1}
  \end{subfigure}%
   \caption{(a) Simulated gain as a function of the over-voltage for an amplification region of $1\,\upmu$m depth for photons.
   \emph{Fit} is the result of a straight-line fit to the data for $\mathit{OV} \ge 3$\,V.
    (b) Relative \emph{rms} widths of the gain distributions. }
  \label{fig:GainSigd1}
 \end{figure}

 The simulated over-voltage dependence of the gain for photons, which is the mean charge in units of elementary charges for the events with Geiger discharges, is shown in Fig.\,\ref{fig:GainSigd1} for an amplification region of $1\,\upmu$m.
  The crosses are the simulations and the diamonds the results of a straight-line fit for $\mathit{OV} \geq 3$\,V.
 The values simulated for photons and for dark counts are statistically compatible, and only the results for photons are shown.

 Different to the results from measurements, significant deviations from linearity are observed at low over-voltages and the intercept with the abscissa is at $\mathit{OV} \approx \,0.5$\,V, which casts doubts on the validity of the simulation.
  The simulations for $d = 0.5\,\upmu$m and for $d = 1.5\,\upmu$m show a similar non-linearity.
   In order to check if, ignoring the currents flowing through $R_q$ and $C_q$ (see Fig.\,\ref{fig:ElScheme}) are responsible for the non-linearity, Eq.\,\ref{eq:dUd} with $R_q = 250$\,k$\Omega$ and $C_q = 2$\,fC has been used for the simulation.
  However, the non-linearity remains.
  Also changing the $x$\,step, $\Delta x$, from 10\,nm to 5\,nm and 20\,nm, the saturation velocity, $v_\mathit{sat}$, from $1\times10^7$\,cm/s to $0.5\times10^7$\,cm/s and $2\times 10^7$\,cm/s, and multiplying the ionization coefficients, $\alpha_h$ and $\alpha_e$, by 0.75 and by 1.5 separately and combined, do not change the \emph{Gain}--\emph{OV} dependence.

 The careful inspection of Fig.\,5b of Ref.\,\cite{Windischhofer:2023}, which shows for $\mathit{OV} = 2$ and 3\,V and $d = 0.5\,\upmu$m the time dependence of the voltage over the avalanche region during the Geiger discharge, shows a similar effect:
  Whereas for $\mathit{OV} = 2$\,V the voltage swing is compatible with $2\cdot \mathit{OV}$, the voltage swing for $\mathit{OV} = 3$\,V is larger than $2\cdot \mathit{OV}$.
 This is in complete agreement with the simulations of this paper.
  It is thus concluded that with respect to the linearity of the mean gain as a function of over-voltage, the simulation fails to describe the experimental data.
 In Fig.\,\ref{fig:SigGaind1} the over-voltage dependence of the relative \emph{rms} width of the gain-distribution is shown.
  For $\mathit{OV} = 0.5$\,V it exceeds 5\,\%, but decreases rapidly and attains values below 0.1\,\% above $\mathit{OV} = 7$\,V.
 The results of the simulations with $d = 0.5\,\upmu$m and $1.5\,\upmu$m are similar.

  \subsection{Simulation with more than one discharge in a pixel for photons}
    \label{subsect:Sim2pixel}

 In this section two separate avalanches generated by two ionizations, separated in time by $\Delta t$, are simulated.
  The question addressed is: Does the Geiger-discharge probability and the gain depend on $\Delta t$?
 Such a dependence could explain the observation reported e.g.\,in Ref.\,\cite{Weitzel:2019} that for sub-nanosecond light pulses the response of SiPMs does not saturate at the value: Number of pixels times the response of a single Geiger discharge.

 The simulation proceeds as discussed in section\,\ref{subsect:Avalanche}.
  The electron-hole pairs are generated at $x = 0.99\,\upmu$m, the first at $t = 0$ and the second at $t = \Delta t$.
 The two avalanches are simulated separately but, for calculating $U_d(t)$, the number of holes and electrons from both avalanches are added.

  \begin{figure}[!ht]
   \centering
   \begin{subfigure}[a]{0.5\textwidth}
    \includegraphics[width=\textwidth]{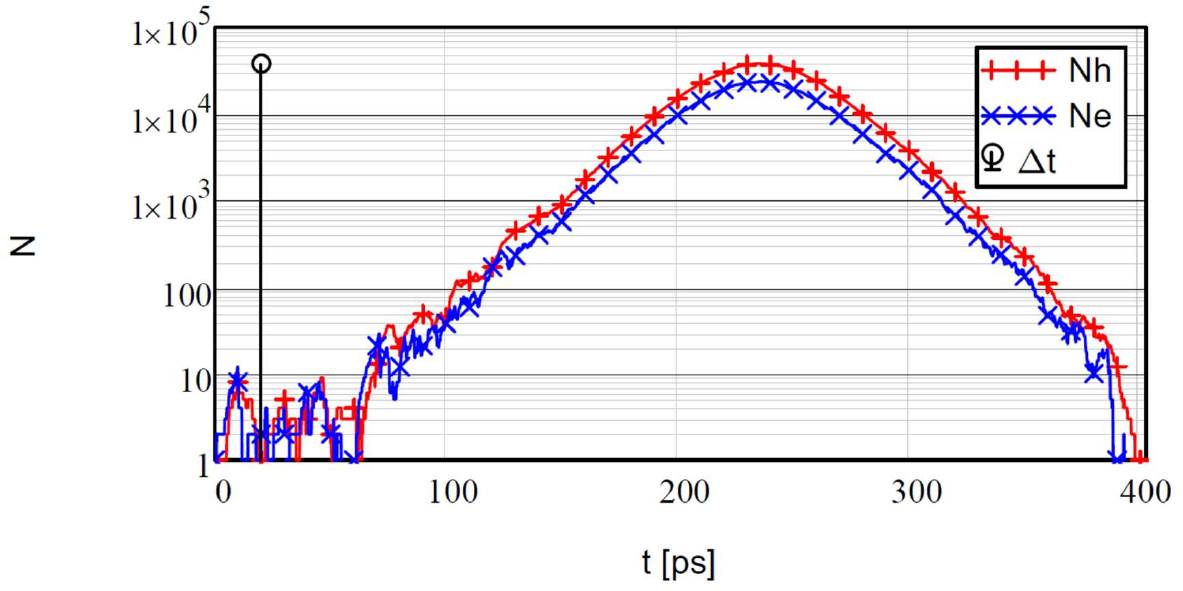}
    \caption{ }
    \label{fig:NtOV1p5bdDt}
   \end{subfigure}%
    ~
   \begin{subfigure}[a]{0.48\textwidth}
    \includegraphics[width=\textwidth]{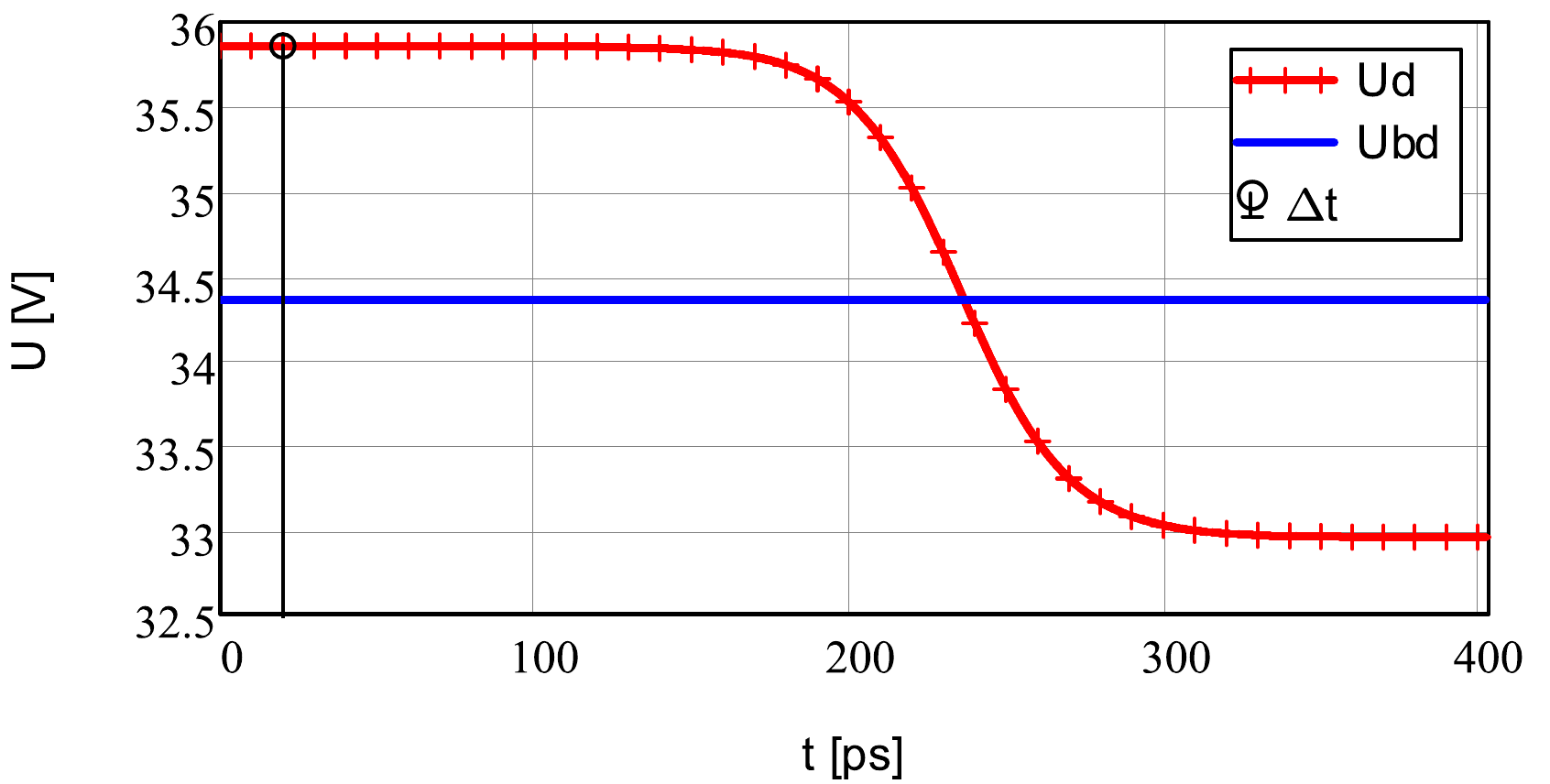}
    \caption{ }
    \label{fig:UdtOV1p5bdDt}
  \end{subfigure}%
   \caption{Simulated time dependence of (a) the number of holes and electrons, and (b) of the voltage over the amplification region for one electron-hole pair generated at $ t = 0$ and one at $t = \Delta t = 20$\,ps.
    The over-voltage is 1.5\,V, the temperature 300\,K and the volume of the amplification region $15 \times 15 \times 1\,\upmu$m$^3$.
   The markers are shown at every hundreds simulated time step.   }
  \label{fig:tOV1p5bdDt}
 \end{figure}

 Fig.\,\ref{fig:tOV1p5bdDt} shows in logarithmic scale the time dependence (a) of the number of holes and electrons, and (b) of $U_d(t)$  for $d = 1\,\upmu$m and $\mathit{OV} = 1.5$\,V for two ionizations at $x = 0.99\,\upmu$m, the first at $t = 0$ and the second at $t = \Delta t = 20$\,ps.
  The distributions are similar to the ones of avalanches induced by a single ionization shown in Fig.\,\ref{fig:NUdt}.

  \begin{figure}[!ht]
   \centering
   \begin{subfigure}[a]{0.5\textwidth}
    \includegraphics[width=\textwidth]{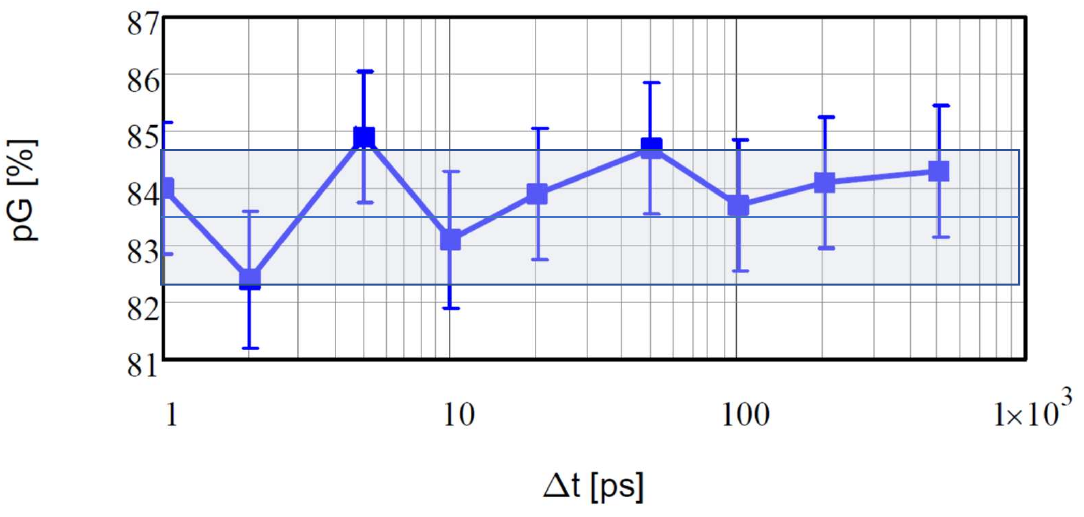}
    \caption{ }
    \label{fig:pG2OV2}
   \end{subfigure}%
    ~
   \begin{subfigure}[a]{0.5\textwidth}
    \includegraphics[width=\textwidth]{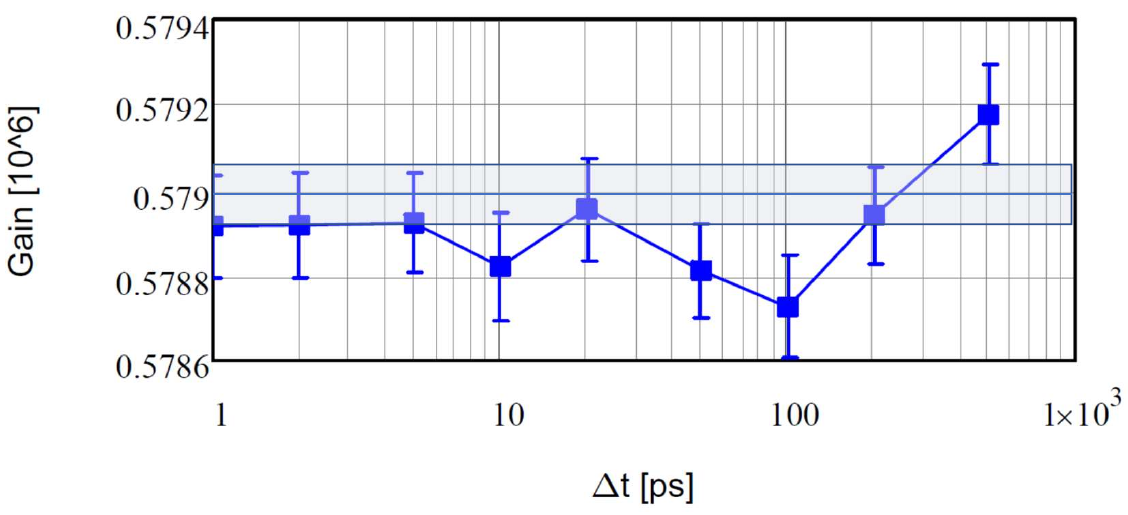}
    \caption{ }
    \label{fig:G2OV2}
  \end{subfigure}%
   \caption{ (a) The simulated Geiger-discharge probability, and (b) the simulated gain for two photons as a function of $\Delta t$, for an amplification region of $1\,\upmu$m depth and a bias voltage of 2\,V.
    The shaded band in (a) is the prediction from \emph{pG} for single photons using Eq.\,\ref{eq:pGn}, and in (b) the gain for single photons. }
  \label{fig:p2OV2}
 \end{figure}

  \begin{figure}[!ht]
   \centering
   \begin{subfigure}[a]{0.5\textwidth}
    \includegraphics[width=\textwidth]{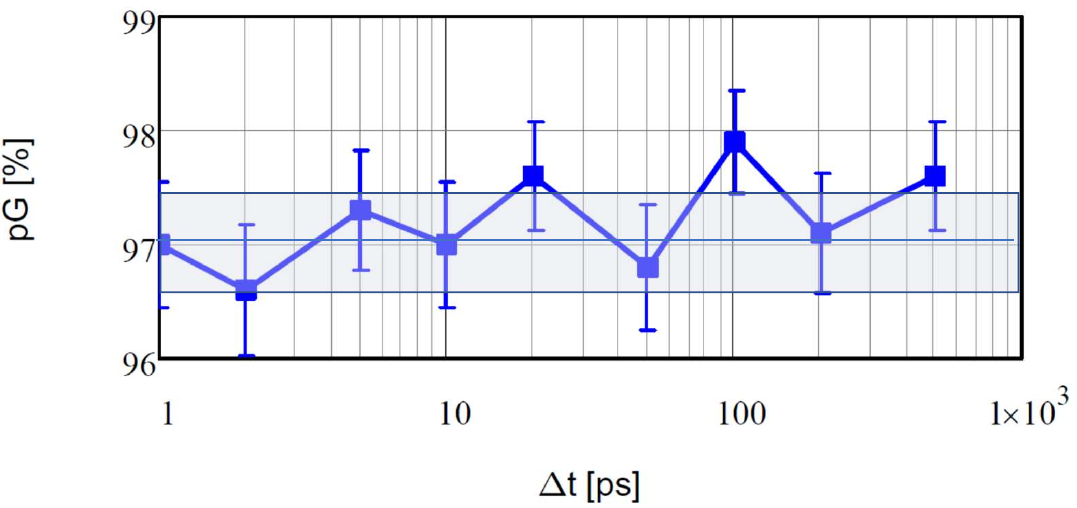}
    \caption{ }
    \label{fig:pG2OV4}
   \end{subfigure}%
    ~
   \begin{subfigure}[a]{0.54\textwidth}
    \includegraphics[width=\textwidth]{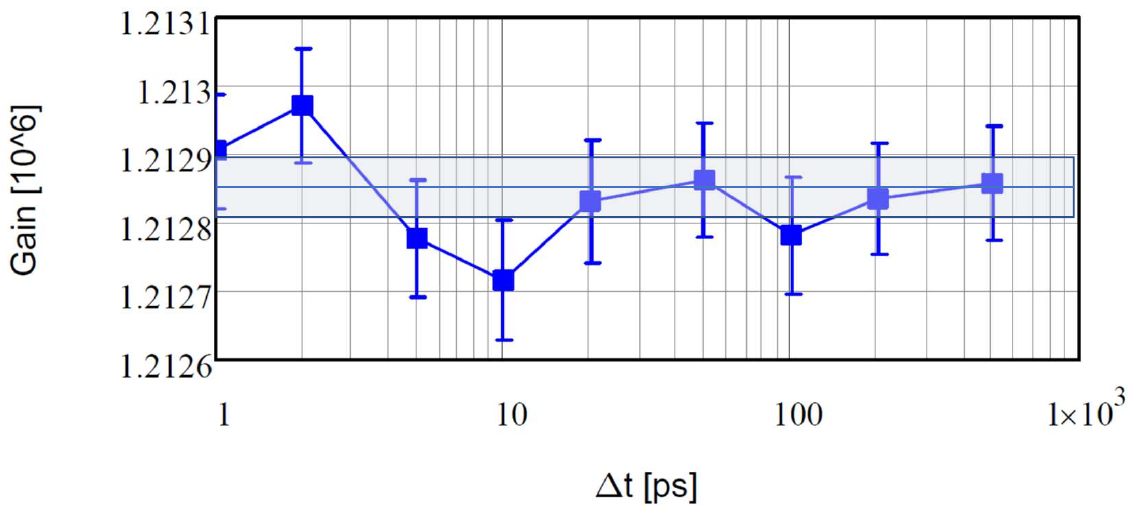}
    \caption{ }
    \label{fig:G2OV4}
  \end{subfigure}%
   \caption{ Same as Fig.\,\ref{fig:p2OV2} for a bias voltage of 4\,V. }
  \label{fig:p2OV4}
 \end{figure}

  Figs\,\ref{fig:p2OV2} and \ref{fig:p2OV4} show as a function of $\Delta t$ the Geiger-discharge probability and the gain for over-voltages of 2\,V and 4\,V, respectively.
   Neither Geiger-discharge probability nor gain depend on $\Delta t$, and thus do not provide an explanation for the observed saturation behaviour of SiPMs exposed to sub-nanosecond light pulses.
  As shown by the shaded bands, the gain for one and for two ionizations by photons is the same.
   The Geiger-discharge probability for $n$ photons, $\mathit{pG}^{(n)}$ can be estimated from the Geiger-discharge probability for 1 photon, $\mathit{pG}^{(1)}$,
 \begin{equation}\label{eq:pGn}
   pG^{(n)} = 1 - \big(1 - pG^{(1)} \big)^n,
 \end{equation}
  where $\big(1 - pG^{(1)} \big)^n$ is the probability that none of the $n$ photons resulted in a Geiger discharge.
   As shown by the shaded bands in Figs\,\ref{fig:pG2OV2} and \ref{fig:pG2OV4}, the simulations agree with this relation.

 \section{Summary and conclusions}
  \label{sect:Conclusions}

 The time development of avalanches in narrow (order $1\,\upmu$m) multiplication regions biased above the breakdown voltage has been simulated with the aim to better understand the functioning of SiPMs.
 The method of Ref.\,\cite{Windischhofer:2023} has been used, making the assumptions of a memory-less electron-hole avalanche for a given dependence of $\alpha_e$ and $\alpha_h$ on the electric field and ignoring space-charge effects in the avalanche.

 It is found that the results differ significantly from the generally accepted understanding of SiPMs:
  The avalanche does not stop when the voltage over the amplification region is close to the breakdown voltage, $U_\mathit{bd}$, but when it is approximately $U_\mathit{bd} - \mathit{OV}$, where the over-voltage, $\mathit{OV}$, is the difference between the bias voltage and $U_\mathit{bd}$.
 As a result the predicted gain is $2\cdot C_d \cdot \mathit{OV}/q_0$ instead of $C_d \cdot \mathit{OV}/q_0$ (elementary charge $q_0$), and no after-pulses are predicted for times $t < \tau \cdot \ln(2)$, when the voltage over the amplification region after a discharge at $t = 0$ starts to exceed $U_\mathit{bd}$; $C_d$ is the capacitance of the multiplication region and $\tau$ the recharging time constant.
  So far this conflict has not been settled by experiments.

  This model \emph{explains} a number of experimental findings:

  \begin{itemize}
    \item The avalanche breakdown happens within less than 1\,ns.
    \item Both dark- and photo-current are proportional to $ \mathit{OV}^a$, where $a$ has only a weak $U_\mathit{bd}$ dependence and is larger for the dark- than for the photo-current.
    \item The \emph{OV} dependence of the Geiger-breakdown probability can be described by $1 - \exp(-\mathit{OV}/U_0)$, where $U_0$ increases with $U_\mathit{bd}$.
        For the geometry simulated, $U_0$ is smaller for photons which generate electron-hole pairs close to the surface, than for dark counts with an approximately uniform distribution over the avalanche region.
  \end{itemize}

 However, there is a major disagreement with experimental results: The dependence of the gain on \emph{OV} is non-linear at small over-voltages, and the extrapolation from the linear region ($\mathit{OV} \gtrsim 3$\,V) reaches gain = 1 at $\mathit{OV}\approx 0.5$\,V.
  This casts doubts on the validity of the simulation with the assumptions made.

  As the model differs drastically from the generally accepted understanding of Geiger breakdown in SiPMs, its predictions should be compared to experimental data.
   In particular the question if the gain is $C_d \cdot \mathit{OV}/q_0$ or twice this value, and if after-pulses appear shortly after the Geiger breakdown or not, should be settled by experiments.
 It is also noted that the two models give very different predictions for Geiger discharges in the same pixel during the pixel recovery time.
   This is in particular relevant for the non-linear regime of high photon intensities and dark-count rates and for after-pulses.










\end{document}